\documentclass[10pt,twocolumn,floatfix]{revtex4-1}
\usepackage{verbatim}
\usepackage{graphicx}
\usepackage{sidecap}
\usepackage{caption}
\usepackage{array}
\usepackage{multirow}
\usepackage{subfigure}
\usepackage{color}
\usepackage[colorlinks,
           bookmarks=true,
           linkcolor=blue,
           urlcolor=blue,
            anchorcolor=black,
            citecolor=blue
            ]{hyperref}
\usepackage{hypcap}
\usepackage{amsmath}

\begin{document}

%\fancyhead[c]{\small Chinese Physics C~~~Vol. 37, No. 1 (2013)
%010201} \fancyfoot[C]{\small 010201-\thepage}

%\footnotetext[0]{Received 14 March 2009}

\title{Elliptic flow of transported and produced protons in Au+Au collisions with the UrQMD model}

\author{Biao Tu}
\affiliation{Key Laboratory of Quarks and Lepton Physics (MOE) and Institute of Particle Physics, Central China Normal University, Wuhan, 430079, China}
\author{Shusu Shi}
\email{shiss@mail.ccnu.edu.cn}
\affiliation{Key Laboratory of Quarks and Lepton Physics (MOE) and Institute of Particle Physics, Central China Normal University, Wuhan, 430079, China}
\author{Feng Liu}
\email{fliu@mail.ccnu.edu.cn}
\affiliation{Key Laboratory of Quarks and Lepton Physics (MOE) and Institute of Particle Physics, Central China Normal University, Wuhan, 430079, China}

%\pacs{25.75.-q}

\begin{abstract}
With the framework of the UrQMD model, by tracing the number of initial quarks in protons, we study the elliptic flow of protons with 3, 2, 1, 0 initial quarks and anti-protons in Au+Au collisions at $\sqrt{s_{NN}}$ = 7.7, 11.5, 39, 200 GeV. The difference of elliptic flow between protons with 2, 1, 0 initial quarks and anti-protons is smaller than 0 or consistent with 0, respectively. The difference of elliptic flow between transported protons (with 3 initial quarks) and anti-protons is larger than 0 at 7.7, 11.5 and 39GeV. It shows a good agreement with the STAR results at 7.7 and  11.5 GeV, but overestimates the STAR results at 39GeV. The yield of transported protons with 3 initial quarks is smaller than that of protons with 2 and 1 initial quarks and the $v_{2}$ of all protons is much smaller than the STAR results. The observation of the difference of elliptic flow between transported protons and anti-protons in the UrQMD model partly explains the $v_{2}$ difference between protons and anti-protons observed in the Beam Energy Scan program in Relativistic Heavy Ion Collider~(RHIC).
\end{abstract}
\maketitle

%\begin{keyword}
%Elliptic flow, UrQMD, Beam Energy Scan
%\end{keyword}

%\begin{pacs}
%25.75.Ld
%\end{pacs}

%\footnotetext[0]{\hspace*{-3mm}\raisebox{0.3ex}{$\scriptstyle\copyright$}2013
%Chinese Physical Society and the Institute of High Energy Physics
%of the Chinese Academy of Sciences and the Institute
%of Modern Physics of the Chinese Academy of Sciences and IOP Publishing Ltd}%

%\begin{multicols}{2}

\section{Introduction}
A strongly interacting hot and dense QCD matter called quark-gluon plasma (QGP) is created in the experiments of high-energy heavy-ion collisions at the Relativistic Heavy Ion Collider (RHIC) and the Large hadron Collider (LHC) ~\cite{Arsene:2004fa,Back:2004je,Adams:2005dq,Adcox:2004mh,Aamodt:2010cz,Aamodt:2010pa}. To understand the properties and phase structure of strongly interacting nuclear matter, the Beam Energy Scan(BES) program involving Au+Au collisions has been carried out at RHIC. Various observables have been measured such as particle production and ratios~\cite{Feng:2018emx,Jin:2018lbk}, Hanbury-Brown-Twiss(HBT) terferometry~\cite{Yang:2016kbz}, moments of the conserved quantities~\cite{Luo:2017faz} and collective flow~\cite{Song:2017wtw}. In this paper, we focus on the second harmonic of collective flow, $v_{2}$. Analyzing the anisotropic flow in nucleus-nucleus collisions is one of the most important directions in studying the properties of created matter~\cite{Ollitrault:1992bk,Sorge:1996pc,Sorge:1998mk}, since it is sensitive to the pressure gradient, degree of freedom, equation of state (EoS)~\cite{Ko:2016ioz,Lu:2016jsv} and degree of thermalization in the early stages of nuclear collisions.

Some interesting phenomena have been observed in heavy-ion collisions of the BES program. The smaller $v_{2}$ of $\overline{p}$, $K^{-}$ and $\pi^{+}$ than those of $p$, $K^{+}$ and $\pi^{-}$, are observed respectively. The $v_{2}$ difference decreases with increasing colliding energy~\cite{Adamczyk:2013gv, Adamczyk:2013gw, Adamczyk:2012gw, Adamczyk:2016gw, sss:2016}. These interesting results were attributed to the different $v_{2}$ of transported and produced quarks during the initial stage of heavy-ion collisions in ref.~\cite{Dunlop:2011cf}. It is argued that the effect results from quark transportation from forward to middle rapidity. The authors assume that the $v_{2}$ of transported quarks is larger than that of produced quarks. Thus, the different numbers of constituent quarks and anti-quarks in the particles and corresponding antiparticles lead to a systematically larger $v_{2}$ of the baryons compared to the anti-baryons. The energy dependence is explained by the increase of nuclear stopping in heavy-ion collisions with decreasing colliding energy~\cite{Li:2018bus}. In ref.~\cite{Burnier:2011bf}, it is suggested that the chiral magnetic effect induced by the strong magnetic field in noncentral collisions could be responsible for the observed difference between the $v_{2}$ of $\pi^{+}$ and $\pi^{-}$.

A calculation~\cite{Xu:2013sta,Song:2012cd} based on the Nambu-Jona-Lasino(NJL) model can also qualitatively explain the difference between $p - \overline{p}$, $\Lambda - \overline{\Lambda}$ and $K^{+} - K^{-}$ by incorporting repulsive potential for quarks and attractive potential for antiquarks, which results in different flow patterns.
The other study~\cite{Xu:2012gf} based on the AMPT model including mean-field potential can also qualitatively explain the difference between the elliptic flow of particles and their corresponding antiparticles.
Because of the more attractive potentials of $\overline{p}$ compared to $p$, smaller $v_{2}$ is obtained for $\overline{p}$. With the attractive $K^{-}$ and repulsive $K^{+}$ potentials, and slightly attractive $\pi^{+}$ and repulsive $\pi^{-}$ potentials, smaller $v_{2}$ are obtained for $K^{-}$ and $\pi^{+}$ than that of $K^{+}$ and $\pi^{-}$.

In this paper, we study the elliptic flows of protons with 3 initial quarks, 2 initial quarks, 1 initial quark, 0 initial quark and anti-protons at BES energies with a ultra relativistic quantum molecular dynamics(UrQMD) model~\cite{Bleicher:1999xi,Bass:1998ca}. The paper is organized as follows: in section~\ref{sec_obs}, the observable is introduced. A brief description of the UrQMD model is given in section~\ref{sec_urqmd}. The results and discussions are presented in section~\ref{sec_results}. Finally, a summary is given in section~\ref{sec_summary}.

\section{Observable}
\label{sec_obs}
The azimuthal anisotropy is one of the most important observables in heavy-ion collisions. In the non-central heavy-ion collisions, the overlap region is an almond shape with
the major axis perpendicular to the reaction plane which is defined by the impact parameter
and the beam direction. As the system evolves, the pressure gradient from the overlapping region of two nuclei in the collisions is the origin of the collective motion component in mid-rapidity. The anisotropy in the coordinate space is transferred to the anisotropy in the momentum space. The anisotropic parameters are defined by the Fourier coefficients of the expansion of the azimuthal distribution~\cite{Voloshin:1994mz, Poskanzer:1998yz} of the produced particles with respect to the reaction plane which can be written as
\begin{eqnarray}
%\frac{dN}{d(\phi-\Psi_{n})} \propto 1 + 2 \sum{v_{n}\mathrm{cos}[n(\phi-\Psi_{n})]},
E\frac{d^{3}N}{dp^{3}} = \frac{1}{2\pi} \frac{d^{2}N}{p_{T}dp_{T}dy}(1 + 2 \sum{v_{n}\mathrm{cos}[n(\phi-\Psi_{\mathrm{RP}})]}),
\end{eqnarray}
where $\phi$ is the azimuthal angle of the particles. $\Psi_{\mathrm{RP}}$ is the reaction plane.
The anisotropic parameter is defined as the $n^{th}$ Fourier coefficient $v_{n}$:
\begin{eqnarray}
v_{n} = \left \langle \mathrm{cos}[n(\phi-\Psi_{\mathrm{RP}})] \right \rangle,
\end{eqnarray}
where $\left \langle \cdots \right \rangle$ is taking the average over all the particles in the sample. The second harmonic coefficient is denoted as elliptic flow $v_{2}$. In the UrQMD model, $\Psi_{\mathrm{RP}}$ is fixed at zero degree.

\section{UrQMD model}
\label{sec_urqmd}
The ultrarelativistic quantum molecular dynamics (UrQMD) model is a microscopic transport model which could simulate the p + p, p + A, and A + A collisions at relativistic energies and describes the time-evolution of a many-body system by using covariant equations of motion. It includes the string excitation and fragmentation, the formation and decay of hadronic resonances, and rescattering of hadrons. At low and intermediate energies, this microscopic transport model focuses on the interactions between known baryon and meson species and their resonances. The excitation and fragmentation of color strings play important roles in the particle production at high energies in the UrQMD model.
The version of the UrQMD model we used in this article is 2.3, and no modification was made to the model itself except for some additional outputs for tracing the particle's origin as explained in ref~\cite{Guo:2012qi}. We marked particles as transported and produced by tracing the number of initial quarks in a particle. In this article, protons with 3, 2, 1 and 0 initial quark are marked as $p$(3 iq), $p$(2 iq), $p$(1 iq) and $p$(0 iq) respectively. There are two special cases. Protons with three initial quarks are treated as transported protons($p$(3 iq)). Protons with zero initial quark are treated as produced protons($p$(0 iq)). Produced protons and anti-protons are both made of three produced quarks, thus they are expected to be similar in many aspects.

\section{Results and discussions}
\label{sec_results}
\begin{figure*}[]
\centering
\includegraphics[width=15cm,clip]{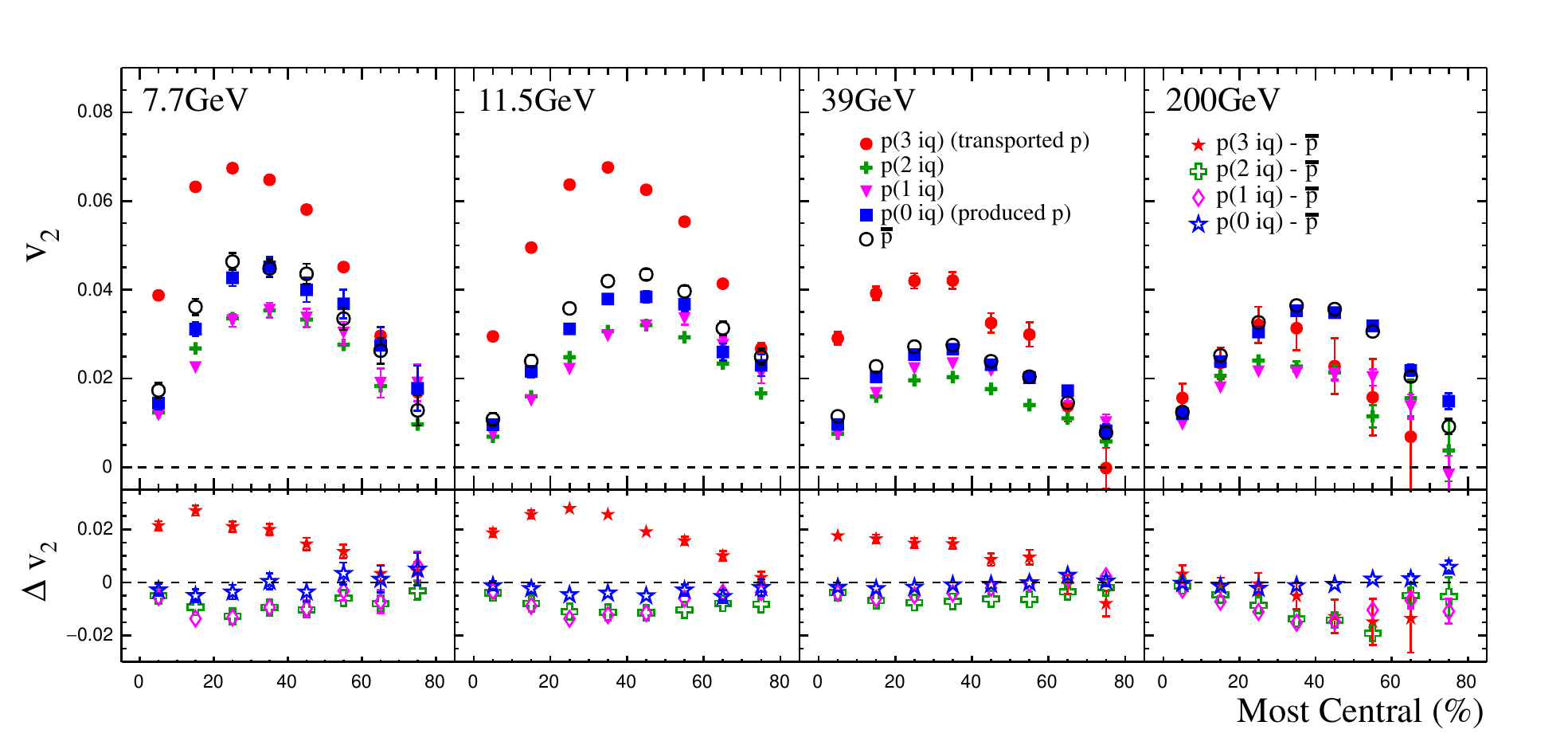}
\caption{(Color online) Upper panel: The elliptic flow of $p$(3 iq), $p$(2 iq), $p$(1 iq), $p$(0 iq) and $\overline{p}$ are plotted as a function of collison centrality in Au+Au collisions at $\sqrt{s_{NN}}$ = 7.7, 11.5, 39, 200~GeV. Lower panel: The difference of $v_{2}$ between $p$(3 iq), $p$(2 iq), $p$(1 iq), $p$(0 iq) and $\overline{p}$ as a function of collision centrality in Au+Au collisions at $\sqrt{s_{NN}}$ = 7.7, 11.5, 39, 200~GeV, respectively.}
\label{v2_cen}       % Give a unique label
\end{figure*}

\begin{figure*}[]
\centering
\includegraphics[width=15cm,clip]{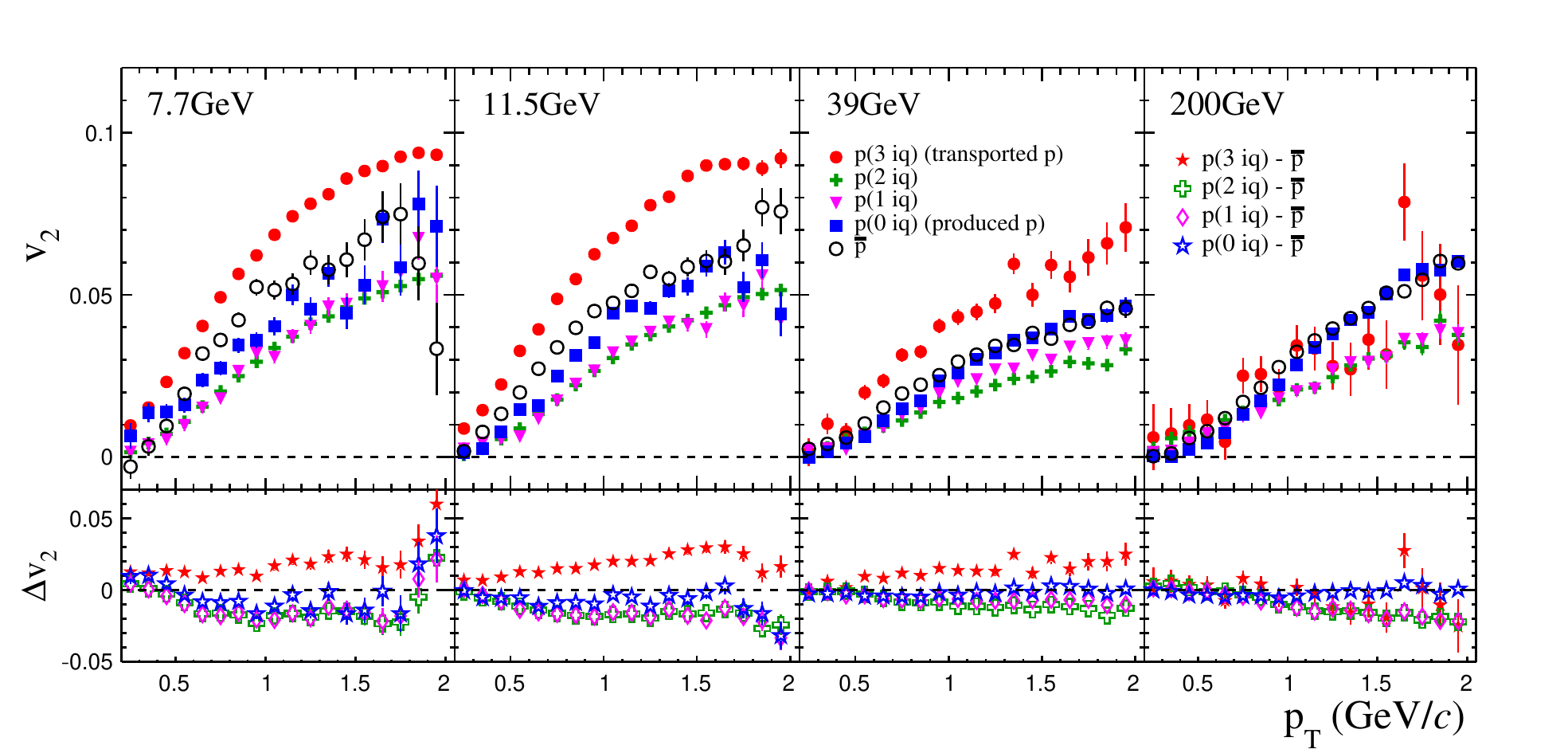}
\caption{(Color online) Upper panel: The elliptic flow of $p$(3 iq), $p$(2 iq), $p$(1 iq), $p$(0 iq) and $\overline{p}$ as a function of the transverse momentum $p_{T}$ for 0-80\% central Au+Au collisions at $\sqrt{s_{NN}}$ = 7.7, 11.5, 39, 200~GeV. The lower panels show the difference of $v_{2}(p_{T})$ between $p$(3 iq), $p$(2 iq), $p$(1 iq), $p$(0 iq) and $\overline{p}$, respectively.}
\label{v2_pt}        % Give a unique label
\end{figure*}

\begin{figure*}[]
\centering
\includegraphics[width=15cm,clip]{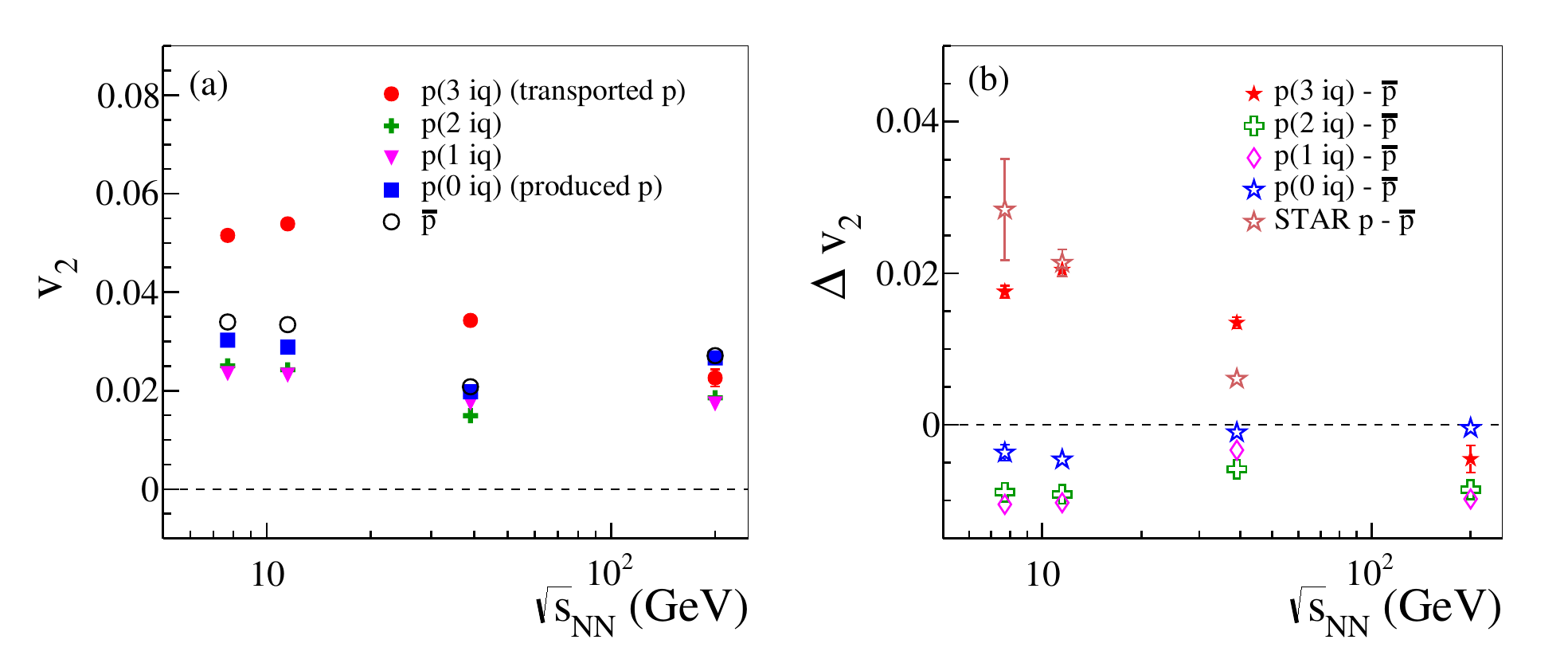}
\caption{(Color online) Panel~(a):~The integrated $v_{2}$ of $p$(3 iq), $p$(2 iq), $p$(1 iq), $p$(0 iq) and $\overline{p}$ as a function of colliding energy for 0-80\% central Au+Au collisions. Panel~(b):~The difference in $v_{2}$ between $p$(3 iq), $p$(2 iq), $p$(1 iq), $p$(0 iq) and $\overline{p}$ as a function of colliding energy for 0-80\% central Au + Au collisions, respectively.}
\label{v2_energy}       % Give a unique label
\end{figure*}
By tracing the number of initial quarks in proton, the $p$(3 iq) account for 17.4\% at 7.7~GeV, 8.3\% at 11.5~GeV, 1.26\% at 39~GeV and 0.36\% at 200~GeV of all protons and anti-protons in the middle rapidity($|\mathrm{Y}|<1$). The $p$(2 iq) account for 80.8\% at 7.7~GeV, 84\% at 11.5~GeV, 52.83\% at 39~GeV and 10.9\% at 200~GeV. The $p$(1 iq) account for 0.91\% at 7.7~GeV, 3\% at 11.5~GeV, 12.46\% at 39~GeV and 14.4\% at 200~GeV. The $p$(0 iq) account for 0.45\% at 7.7~GeV, 2.4\% at 11.5~GeV, 15.4\% at 39~GeV and 39.1\% at 200~GeV. The $\overline{p}$ account for 0.44\% at 7.7~GeV, 2.3\% at 11.5~GeV, 18.05\% at 39~GeV and 35.24\% at 200~GeV. The upper panel of Figure~\ref{v2_cen} shows the elliptic flow of $p$(3 iq), $p$(2 iq), $p$(1 iq), $p$(0 iq) and $\overline{p}$ within $0.2 < p_{T} < 2.0~\mathrm{GeV}/c$ as a function of collision centrality in Au+Au collisions at various colliding energies. One can find that $v_{2}$ shows strong centrality dependence since it is mainly driven by the initial spatial eccentricity. The lower panel shows the difference of $v_{2}$ between $p$(3 iq), $p$(2 iq), $p$(1 iq), $p$(0 iq) and $\overline{p}$, respectively. The difference of $v_{2}$ between $p$(0 iq) (produced) and $\overline{p}$ does not show clearly centrality dependence, and is approximately consistent with 0. As we mentioned above, the $p$(0 iq) should be similar to $\overline{p}$ in many aspects. They are all made up by produced quarks. The $p$(0 iq) and $\overline{p}$ both are produced at the early stage of the system that energy density is relatively large in low colliding energies. Both $p$(0 iq) and $\overline{p}$ experience the full evolution of the system which lead to a similar magnitude of $v_{2}$. Larger elliptic flow of $p$(3 iq) than that of $p$(0 iq) and $\overline{p}$ is observed. It suggests the $v_{2}$ of $p$(3 iq) transported from forward rapidity to mid-rapidity due to nuclear stopping effect is different from the $v_{2}$ of $p$(0 iq) and $\overline{p}$. These results indicate that the elliptic flow of transported quarks is larger than that of produced quarks. The transported quarks which have been transported over a large rapidity suffer more scatterings than the produced quarks. It develops a larger $v_{2}$ of transported quarks than that of produced quarks, thus can lead to a larger $v_{2}$ of $p$(3 iq) than that of $p$(0 iq).
The difference of $v_{2}$ between $p$(3 iq) and $\overline{p}$ shows a strong centrality dependence. Larger difference is observed in middle central collisions than that in most central and peripheral collisions at 7.7, 11.5 and 39 GeV. The $p$(3 iq) experiences the whole process that the initial geometry eccentricity is transformed into anisotropy in the momentum space, whereas the $p$(0 iq) may only partly experiences the process. The combination of baryon stopping effect and scatterings makes the difference of $v_{2}$ between the transported protons and anti-protons largest in mid-central collisions. No significant centrality dependence is observed at 200~GeV due to the small difference of $v_{2}$ between $p$(3 iq) and $\overline{p}$.

The $v_{2}$ of $p$(2 iq) and $p$(1 iq) show similar centrality dependence with that of $v_{2}$ of $p$(0 iq) and $\overline{p}$, but are systematically lower than the $v_{2}$ of $p$(0 iq) and $\overline{p}$. The $p$(0 iq), $p$(1 iq), $p$(2 iq) and $\overline{p}$ are produced at the same time at early stage based on a string-excitation scheme~\cite{Guo:2012qi}, but part of $p$(1 iq) and $p$(2 iq) are produced by the decay of unstable baryons. It means the formation time of $p$(0 iq) and $\overline{p}$ should be earlier than that of $p$(1 iq) and $p$(2 iq). So $p$(1 iq) and $p$(2 iq) suffer less interactions than $p$(0 iq) and $\overline{p}$. The $v_{2}$ of $p$(1 iq) and $p$(2 iq) is smaller than that of $p$(0 iq) and $\overline{p}$. Thus, in the UrQMD model, the $v_{2}$ of inclusive $p$ is slightly lower than or consistent with the $v_{2}$ of $\overline{p}$ depending on the collision energy which is consistent with the results in ref~\cite{Li:2016szd}.

The upper panel of Figure~\ref{v2_pt} shows the elliptic flow of $p$(3 iq), $p$(2 iq), $p$(1 iq), $p$(0 iq) and $\overline{p}$ as a function of transverse momentum $p_{T}$ in 0-80\% Au+Au collisions at $\sqrt{s_{NN}}$ = 7.7, 11.5, 39, 200~GeV. The lower panel shows the difference of $v_{2}$ between $p$(3 iq), $p$(2 iq), $p$(1 iq), $p$(0 iq) and $\overline{p}$, respectively. The difference of $v_{2}$ between $p$(0 iq) and $\overline{p}$ does not show a clearly $p_{T}$ dependence and is almost consistent with 0 except 7.7 GeV. But the difference of $v_{2}$ between $p$(3 iq) and $\overline{p}$ shows a weak $p_{T}$ dependence. The splitting of $v_{2}$ between $p$(3 iq) and $p$(0 iq) may be due to the stronger flow of transported quarks which experience more interactions than produced quarks. This phenomenon is consistent with the study in ref~\cite{Dunlop:2011cf}, by assuming the $v_{2}$ of transported quarks is larger than that of produced quarks, and it results in a splitting of $v_{2}$ between protons and anti-protons. The $v_{2}$ of $p$(2 iq) and $p$(1 iq) increase with $p_{T}$, but are systematically smaller than that of $p$(3 iq), $p$(0 iq) and $\overline{p}$.

To compared with the STAR results, we present the $p_{T}$ integrated $v_{2}$ of $p$(3 iq), $p$(2 iq), $p$(1 iq), $p$(0 iq) and $\overline{p}$ within $0.2 < p_{T} < 2.0~\mathrm{GeV}/c$ as a function of colliding energy. In Figure~\ref{v2_energy}, panel~(a) shows the integrated elliptic flow $v_{2}$ of $p$(3 iq), $p$(2 iq), $p$(1 iq), $p$(0 iq) and $\overline{p}$ as a function of colliding energy in Au+Au collisions. Panel~(b) shows the difference in $v_{2}$ from STAR measurements and UrQMD model as function of colliding energy in Au+Au collisions. The $v_{2}$ of $p$(3 iq) is systematically larger than that of $p$(0 iq) and $\overline{p}$. Thus the $v_{2}$ difference between $p$(3 iq) and $\overline{p}$ is larger than 0. The $v_{2}$ difference between $p$(0 iq) and $\overline{p}$ is slightly smaller than 0 or consistent with 0 depending on the colliding energy. The $v_{2}$ of $p$(2 iq) and $p$(1 iq) are systematically smaller than that of $p$(3 iq), $p$(0 iq) and $\overline{p}$. So the difference of $v_{2}$ between $p$(2 iq)/$p$(1 iq) and $\overline{p}$ are smaller than 0. The difference of $v_2$ between the $p$(3 iq) and $\overline{p}$ show a similar energy dependence compared with the STAR results.
Our results of $p$(3 iq) - $\overline{p}$ show a good agreement with the STAR results below 11.5~GeV. At 39~GeV, the $v_{2}$ difference of $p$(3 iq) and $\overline{p}$ are not consistent with the STAR results quantitatively. The yield of $p$(3 iq) is relatively smaller than that of $p$(2 iq) and $p$(1 iq), the $v_{2}$ of all protons is much smaller than that in the STAR results. This study indicates that the $v_{2}$ difference of STAR measurements may be partly due to the $v_{2}$ difference between $p$(3 iq) and $\overline{p}$.  But it can not explain the STAR results. In principle, the yield of $p$(3 iq) dominates the yield of protons at low energies. The magnitude of $v_{2}$ difference of $p$(3 iq) - $\overline{p}$ being consistent between STAR data and UrQMD model in Au+Au collisions at 7.7 and 11.5~GeV suggests that the hadronic interactions are dominant in these collision energies. The derivation of $v_{2}$ difference between STAR results and our calculations at 39~GeV indicates that the partonic interactions are also important to build up $v_{2}$ at high energies. Additionally, with the energy increasing the fraction of $p$(3 iq) relative to inclusive protons decreases can also lead to such a derivation.

\section{Summary}
\label{sec_summary}
In summary, by tracing the number of initial quarks in the UrQMD model, the $p$(3 iq), $p$(2 iq), $p$(1 iq), $p$(0 iq) can be distinguished. It provides a way to study the elliptic flow of transported protons and produced protons. We found that the elliptic flow of produced protons shows similar dependence on collision centrality, transverse momentum and colliding energy with anti-protons. The possible explanation is produced protons and anti-protons are both made up of produced quarks. At the same time, the produced protons and anti-protons can be only produced at the early stage in the hadronic evolution of the system. Both of them experience the similar magnitude of interactions in the system which lead to similar $v_{2}$. For the transported protons, the elliptic flow is systematically larger than that of anti-proton as a function of collision centrality, transverse momentum and colliding energy. This can be explained as following:
quarks transported from forward rapidity to mid-rapidity by the baryon stopping effect gain larger $v_{2}$ than produced quarks.
The $v_{2}$ of $p$(2 iq) and $p$(1 iq) are systematically smaller than that of transported protons, produced protons and anti-protons. Our results with the UrQMD model indicate that the splitting of $v_{2}$ for protons may partly arise from the difference of $v_{2}$ between transported quarks and produced quarks. The difference of $v_{2}$ between transported protons with 3 initial quarks and anti-protons show a good quantitative agreement with that between protons and anti-protons in STAR measurements at low energies~(7.7 and 11.5~GeV), but large deviation at high energies~($\ge$ 39~GeV). It suggests the hardonic interactions are dominant in collisions at 7.7 and 11.5~GeV. Without partonic interactions in the UrQMD model, it is problematic to reproduce the $v_{2}$ at higher energy~($\ge$ 39~GeV). On the other hand, the fraction of transported protons relative to inclusive protons may also attribute the derivation between STAR data and the UrQMD model.

\vspace{-1mm}
\centerline{\rule{80mm}{0.1pt}}
\vspace{2mm}

\begin{acknowledgments}
This work is supported in part by the MoST of China 973-Project No. 2015CB856901, National Natural Science Foundation of China under Grants No. 11890711 and self-determined research funds of CCNU from the colleges¡¯ basic research and operation of MOE under Grant No. CCNU18TS031.
\end{acknowledgments}

%\end{multicols}
\clearpage

%\end{CJK*}
\end{document}